\documentclass[11pt,onecolumn]{IEEEtran}
\usepackage{amssymb} 

\begin{document}

\title{A Decision-Optimization Approach \\to Quantum Mechanics and Game Theory}

\author{{\small Xiaofei Huang \\ 
	Cccent Inc., Foster City, CA 94404\\HuangZeng@yahoo.com}}

\maketitle

\begin{abstract}
The fundamental laws of quantum world upsets the logical foundation of classic physics.
They are completely counter-intuitive with many bizarre behaviors. 
However, this paper shows 
	that they may make sense from the perspective of a general decision-optimization principle for cooperation. 
This principle also offers a generalization of Nash equilibrium, a key concept in game theory,
	for better payoffs and stability of game playing.
\end{abstract}


In physics, discovering more fundamental laws that can be used to derive some existing laws 
	usually gives us a better understanding about nature, 
	e.g., the discovery of Newton's laws of motion and gravitation for explaining Kepler's laws of planetary motion.
This paper shows that a global optimization algorithm can be used to derive the time-independent Schr\"{o}dinger equation,
	one of the most important equations in quantum mechanics.
Such a connection suggests that the laws of quantum mechanics might embed a global optimization process 
	such that a dissipative quantum system, caused by the interactions with its environment,
	tends to evolve towards its ground state of the global energy minimum.
Contrary to that, the same system under the laws of classic physics
	tends to get stuck at one local minimum energy state or another.
The former is more deterministic and predictable than the latter
	in terms of the final result (no dice playing) because an energy function usually has an enormous number of local minima but only one global minimum.
In addition to a computational approach to quantum mechanics,
	the global optimization algorithm also offers an intuitive, direct mechanism 
	to compute the stationary states of a quantum system starting from any random initial states.

The global optimization algorithm is defined by the cooperative optimization theory~\cite{HuangBookCCO}.
The theory is for a mathematical understanding of ubiquitous cooperative behaviors in nature and translating them into optimization algorithms.
Specifically, given a society with $n$ persons, assume that 
	the objective of person $i$ is described as minimizing a function $E_i(x)$.
A simple form of cooperative optimization is defined as an iterative computation 
	of each person's expected returns described by a function $\Psi_i (x_i, t)$ as follows:
\begin{equation}
\Psi_i (x_i, t) = \sum_{\sim x_i} \left( e^{-E_i(x)/\hbar} \prod_{j \not= i} p_j(x_j, t-1) \right),\quad \mbox{for $i=1,2,\ldots,n$}  \ , 
\label{cooperative_optimization_general3}
\end{equation}
where $\sum_{\sim x_i}$ stands for the summation over all variables except $x_i$.
$\hbar$ is a constant of a small positive value.
$p_i(x_i, t)$ is a probability-like function for picking action $x_i$ proportional to $\left(\Psi_i(x_i,t)\right)^{\alpha} (\alpha > 0)$, i.e.,
\[ p_i(x_i, t) = \left(\Psi_i(x_i,t)\right)^{\alpha} / Z_i(t), \quad \mbox{where } Z_i(t) = \sum_{x_i} \left(\Psi_i(x_i,t)\right)^{\alpha} \mbox{is a normalization factor.} \]

The larger the parameter $\alpha$, the more aggressive each person is at minimizing his own objective function $E_i(x)$.
At the same time, the game tends to have more equilibrium points.
However, the chance for the society to reach the social (global) optimum
	is only peaked at a certain value of $\alpha$, neither too large nor too small.
In this case, each person in the game compromises his best action by accepting sub-optimal actions to some degree,
	different from the case when $\alpha \rightarrow \infty$ where only the best action is accepted.
	
In particular, when $\alpha = 2$, 
	the simple general form~(\ref{cooperative_optimization_general3})  in a continuous-time version is
\begin{equation}
-\hbar \frac{\partial \psi_i (x_i, t)}{\partial t} = \frac{1}{Z_i (t)} e_i(x_i) \psi_i (x_i, t), \quad \mbox{where } e_i(x_i)=\sum_{\sim x_i} \left( E_i (x) \prod_{j \not= i} |\psi_j(x_j, t)|^2  \right) \ . 
\label{cooperative_optimization_general3.2}
\end{equation}

Following the notation from physics, denote $\psi_i (x_i, t)$ as a vector $ \mid \psi_i (t) \rangle$.
Let $H_i$ be a diagonal matrix with diagonal elements as $e_i(x_i)$.
Then the equation~(\ref{cooperative_optimization_general3.2}) becomes
\begin{equation}
-\hbar \frac{d}{d t} \mid \psi_i (t) \rangle = \frac{1}{Z_i (t)} H_i \mid \psi_i (t) \rangle \ . 
\label{cooperative_optimization_general3.3}
\end{equation}
The above equation can be further generalized with a hermitian matrix $H_i$.

The expected return function $\psi_i (x_i, t)$ is also called a wavefunction in physics.
It is important to note that the equation~(\ref{cooperative_optimization_general3.3}) 
	is the dual equation of the Schr\"{o}dinger equation,
where $-1$ is replaced by the imaginary unit $i$ and the normalization factor $Z_i(t)$ is not required
	since the equation is unitary.
When the dynamic equation~(\ref{cooperative_optimization_general3.3}) 
	reachs a stationary point (equilibrium),
	the equation becomes the time-independent Schr\"{o}dinger equation:
\[ \lambda_i \mid \psi_i (x_i, t) \rangle = H_i \mid \psi_i (x_i, t) \rangle \ , \]
where $\lambda_i$ can only be any one of the eigenvalues of $H_i$.

In summary, when the global optimization algorithm (with $\alpha=2$) in a continuous-time version
	reaches any equilibrium point, 
	it falls back to the time-independent Schr\"{o}dinger equation.
The author is convinced that the algorithm is general 
	and can be applied to solve optimization problems from different fields,
	such as game theory described below.
    

Let $u_i (x) = e^{-E_i(x)/\hbar}$ be the utility function for the person $i$.
In this case, the person $i$ tries to maximize his utility function $u_i(x)$ instead of 
	minimizing his objective function $E_i(x)$.
Both tasks are fully equivalent to each other.
In this case, (\ref{cooperative_optimization_general3}) becomes as
\begin{equation}
\Psi_i (x_i, t) = \sum_{\sim x_i} \left( u_i(x) \prod_{j \not= i} p_j(x_j, t-1) \right),~~\mbox{where } p_i(x_i, t) = \left(\Psi_i(x_i,t)\right)^{\alpha} / \sum_{x_i} \left(\Psi_i(x_i,t)\right)^{\alpha}. 
\label{cooperative_optimization_general4}
\end{equation}
Based on Brouwer fixed point theorem, an equilibrium point always exists for the above set of equations for any value of $\alpha$. 
This point is also an $\epsilon$-approximate Nash equilibrium~\cite{nash50},
	where $\epsilon$ is inversely proportional to $\alpha$.
An $\epsilon$-approximate Nash equilibrium is a strategy profile such that no other strategy can improve
	the payoff by more than the value $\epsilon$.
In particular, a Nash equilibrium can be viewed as an $0$-approximate one.

When $\alpha \rightarrow \infty$, 
	the approximation error $\epsilon$ is arbitrarily close to zero. 
In this case, each player in a game only accepts the best action that gives the highest payoff for the player.
That is the fundamental principle of rational decision making in classic game theory. 
The logical justification of this principle seems obvious
	which shapes the definition of Nash equilibrium more than 50 years ago. 
However, the understanding of the global optimization algorithm~(\ref{cooperative_optimization_general3})
	shows that the optimality of this principle is conditional.
Often times, compromising it by accepting sub-optimal actions can improve both the overall payoff (equivalently the average individual payoff)
	and the stability of game playing.

In summary, the optimal decision of each player in the classic sense 
	may not lead to a good payoff for the player.
At a Nash equilibrium, if every player gives away some payoff 
	by accepting sub-optimal actions to some degree, 
	each of them may actually receive a better, rather than worse return at a new equilibrium point than the original one,
	which may be counter-intuitive.
The experimental results from the prisoner's dilemma to computer simulated societies
	show that compromising each individual's optimal decision
	can improve the overall payoff (many times everyone's payoff) and social stability.
This study suggests that, for the benefit of everyone in a society (or a financial market), 
	the pursuit of maximal payoff by each individual should be controlled at some level
	either by voluntary good citizenship or by imposed regulations.
	
In conclusion, decision compromising is a general optimization principle for cooperation. 
It reveals a global optimization approach to understand quantum mechanics.
It also offers a fundamental principle for improving social stability and individual payoffs 
	over the classic profit-maximization principle.


\begin{thebibliography}{8}

\bibitem{HuangBookCCO}
X.~Huang, ``Cooperative optimization for solving large scale combinatorial
  problems,'' in Theory and Algorithms for Cooperative Systems, 
  Series on Computers and Operations Research, World Scientific, 2004, pp. 117--156.

\bibitem{Tegmark01}
M.~Tegmark and J.~A. Wheeler, ``100 years of quantum mysteries,''
  \emph{Scientific American}, pp. 68--75, February 2001.

\bibitem{Seife05}
C.~Seife, ``Do deeper principles underlie quantum uncertainty and
  nonlocality?'' \emph{Science}, vol. 309, no. 5731, p.~98, July 2005.

\bibitem{nash50}
Nash, J.F.:
\newblock Equilibrium points in n-player games.
\newblock In: Proceedings of the National Academy of Sciences of the United
  States of America. Volume 36(1). (1950)  48--49

\end{thebibliography}
\end{document}